\documentclass[twocolumn,aps,floatfix,superscriptaddress,showpacs]{revtex4-1}
\usepackage{amsmath,amssymb}
\usepackage{subfigure}
\usepackage{graphicx}
\usepackage{setspace}
\usepackage{color}
\usepackage{xspace}
\usepackage{ucs}
\usepackage[absolute]{textpos}

\newcommand{\tlead}{\ensuremath{t_\text{lead}}\xspace}
\newcommand{\tstep}{\ensuremath{t_\text{step}}\xspace}
\newcommand{\emv}[2]{\ensuremath{\mathbf{s}_{#1}^{#2}\xspace}}
\newcommand{\mx}{\ensuremath{\mathbf{x}}\xspace}
\newcommand{\mxc}[2]{\ensuremath{x_{#1}^{#2}\xspace}}
\newcommand{\sit}[2]{\ensuremath{\mathbf{s}_{#1}^{#2}\xspace}}
\newcommand{\mxct}[2]{\ensuremath{\tilde{x}_{#1}^{#2}(m_t,m_s)\xspace}}

\begin{document}

\title{Data-driven prediction and prevention of extreme events in a spatially extended excitable system}

\author{Stephan Bialonski}
\email{bialonski@gmx.net}
\affiliation{Max Planck Institute for the Physics of Complex Systems, N\"othnitzer Stra\ss{}e~38, 01187~Dresden, Germany}

\author{Gerrit Ansmann}
\affiliation{Department of Epileptology, University of Bonn, Sigmund-Freud-Stra\ss{}e~25, 53105~Bonn, Germany}
\affiliation{Helmholtz Institute for Radiation and Nuclear Physics, University of Bonn, Nussallee~14--16, 53115~Bonn, Germany}
\affiliation {Interdisciplinary Center for Complex Systems, University of Bonn, Br\"uhler Stra\ss{}e~7, 53175~Bonn, Germany}

\author{Holger Kantz}
\affiliation{Max Planck Institute for the Physics of Complex Systems, N\"othnitzer Stra\ss{}e~38, 01187~Dresden, Germany}

\begin{abstract}
Extreme events occur in many spatially extended dynamical systems, often devastatingly affecting human life which makes their reliable prediction and efficient prevention highly desirable.
We study the prediction and prevention of extreme events in a spatially extended system, a system of coupled FitzHugh--Nagumo units, in which extreme events occur in a spatially and temporally irregular way.
Mimicking typical constraints faced in field studies, we assume not to know the governing equations of motion and to be able to observe only a subset of all phase-space variables for a limited period of time.
Based on reconstructing the local dynamics from data and despite being challenged by the rareness of events, we are able to predict extreme events remarkably well.
With small, rare, and spatiotemporally localized perturbations which are guided by our predictions, we are able to completely suppress extreme events in this system.
\end{abstract}


\maketitle

\begin{textblock*}{20cm}(3cm,27cm)
 Published as Phys. Rev. E \textbf{92}, 042910 (2015). Copyright 2015 by the American Physical Society.
\end{textblock*}

\section{Introduction}

The dynamics of very different systems can exhibit rare, recurrent, and strong deviations from the regular behavior.
Since such extreme events can often severely impact human life, they are intensively studied~\cite{Albeverio2006,Ghil2011,Sornette2006,Bunde2002} in physics and mathematics as well as in diverse scientific disciplines such as atmospheric sciences (e.g., hurricanes, floods, droughts)~\cite{Field2012,Peterson2013}, oceanography (rogue ocean waves~\cite{Dysthe2008}), geophysics (earthquakes, volcanic eruptions), economics (stock market crashes~\cite{Sornette2003}), engineering (outages in infrastructure~\cite{Dobson2007}), biology (harmful algal blooms~\cite{Anderson2012a}) and medical science (epidemics, heart attacks, epileptic seizures~\cite{Engel2007}). Many of these extreme events occur in spatially extended systems in which they start localized and later propagate.
A successful and reliable forecast of individual extreme events is highly desirable since it could not only provide a warning time during which precautions could be taken but may also, depending on the system dynamics, allow for appropriate countermeasures.

In field studies, we typically cannot record the temporal evolution of all degrees of freedom of such systems, we typically do not know the exact equations of motion, and our periods of measurement are finite during which extreme events, which are intrinsically rare~\cite{Kantz2006}, are to be observed.
This poses a challenge for any prediction attempt.
Thus, research has often focused on statistical aspects (supported by extreme value theory~\cite{Coles2001}) to determine the relative frequency of extreme events, on modeling studies, which can provide a better understanding of the underlying mechanisms leading to extreme events and their predictability, as well as on model-assisted prediction, in which model assumptions and data assimilation techniques deal with our incomplete knowledge about the system dynamics.

In deterministic systems, the prediction of extreme events can profit from the fact that the current state of the system uniquely determines its future state, and is limited in chaotic systems (among other factors~\cite{Siegert2015}) by the sensitive dependence on initial conditions (Butterfly Effect).
In field studies, nonlinear time-series-analysis techniques~\cite{Ott1994b,Kantz2003} allow one, under quite general assumptions, to construct an embedding space out of empirical data that is topologically equivalent to the often unknown state space of the system.
The prediction of extreme events can then be based on the identification of similar states \emph{(analogues)} in the observed past of the dynamics.
Approaches based on a reconstructed space have been successfully pursued for low-dimensional systems (e.g., to detect early warning signs of runaway initiations in chemical reactors~\cite{Zaldivar2005}) and usually exploit one time series of a single observable only.
However, in spatially extended systems, the prediction of extreme events based on such a scheme is a nontrivial challenge, particularly if extreme events emerge in different locations or if the system is heterogeneous.

\begin{figure*}
\begin{center}
 \includegraphics[width=\textwidth]{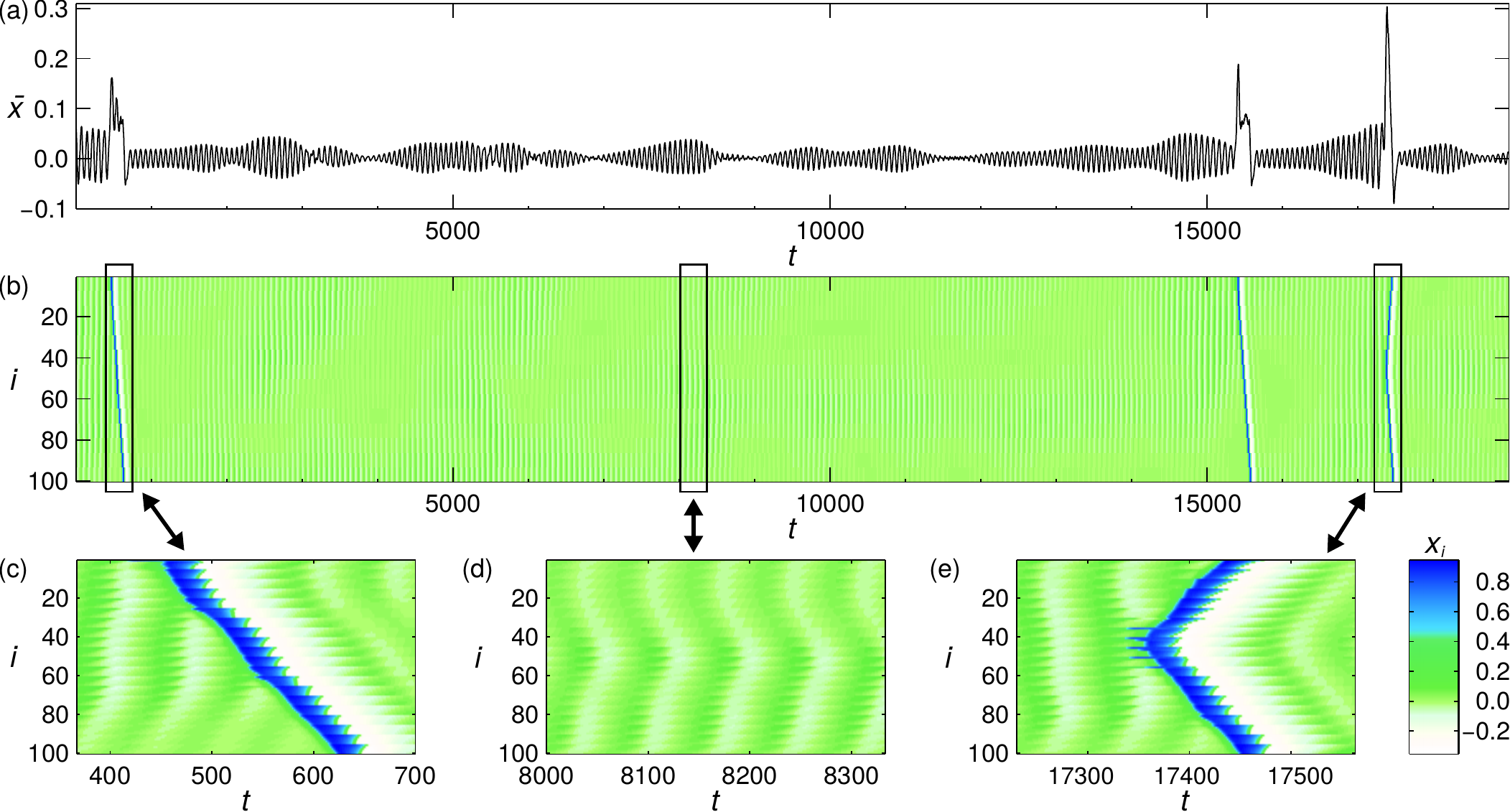}
\end{center}
\caption{
(Color online) (b) Exemplary evolution of the dynamical variables $x_i$ (color coded) for all 100 units as well as (a) the evolution of the corresponding average value $\overline{x}(t) = N^{-1} \sum_{i=1}^{N} x_i(t)$.
Panels (c), (d) and (e) show enlarged segments of (b).
}
\label{fig:1}
\end{figure*}

In this paper, we address this problem by adapting the concept of local embedding spaces~\cite{Parlitz2000} in order to locally predict extreme events that occur spontaneously and in a temporally and spatially irregular way.
Mimicking the situation in field studies, we assume that only some but not all degrees of freedom of the system have been observed, that our period of measurement is finite, and that the equations of motion are unknown (except for the coupling topology). 
Based on our predictions, we prevent extreme events via rare, small, and spatiotemporally localized interventions.
We demonstrate our approach in a case study by predicting and preventing extreme events that occur in an excitable, spatially extended, reaction--diffusion-type dynamics.
Examples of real-world phenomena which may be in this class include epileptic seizures~\cite{Engel2007}, harmful algal blooms~\cite{Medvinsky2002,Anderson2012a}, and cardiac arrhythmia~\cite{Qu2014}.

We show how local embedding spaces can be defined and iterative predictions can be made (section~\ref{sec:embspaces}) for exemplary time series generated by a model (section~\ref{sec:system}).
We observe this method to predict not only the occurrence but also the spatial propagation and termination of extreme events (i.e.\ their full life cycle) remarkably well (section~\ref{sec:prediction}), which we statistically evaluate (section~\ref{sec:statistics}) and exploit to efficiently prevent extreme events via minimal interventions (section~\ref{sec:prevention}).
Finally, we discuss findings and potential ways to further improve prediction performance (section~\ref{sec:discussion}).

\section{A spatially extended system with extreme events}
\label{sec:system}
We consider a spatially extended system whose dynamics shows extreme events, i.e.\ rare and recurrent events that deviate from the usual dynamics.
From this system, time series are obtained which are used in the subsequent sections in order to demonstrate our time series based approach towards the prediction of extreme events.
Our system is inspired by models investigated in recent studies of networks of coupled FitzHugh--Nagumo units \cite{Ansmann2013,Karnatak2014}.
The FitzHugh--Nagumo unit, also known as Bonhoeffer--van der Pol oscillator \cite{VanDerPol1928,Bonhoeffer1948,FitzHugh1961,Nagumo1962}, has two degrees of freedom (a voltage-like variable~$x$ and a recovery variable~$y$), can show a variety of dynamics depending on parameter values~\cite{Rocsoreanu2000}, and can display a characteristic pulse of activity in response to an external stimulus---the hallmark of excitability~\cite{Izhikevich2007}.

We consider a network of $N$ diffusively coupled FitzHugh--Nagumo units in which the dynamics of the $i$th unit is given by
\begin{equation}
 \begin{aligned}
 \dot{x}_i &= x_i (a-x_i)(x_i-1) - y_i + k\sum_{j=1}^{N}A_{ij}(x_j-x_i),\\
 \dot{y}_i &= b_i x_i - c y_i,
 \end{aligned}
 \label{eq:system}
\end{equation}
where $A$ denotes the adjacency matrix of a network, $k=0.0128$ is the coupling strength, $a = -0.03$, $c = 0.02$, and~$b_i$ are internal parameters, and $i\in\{1,\ldots,N\}$.
\emph{Parameter heterogeneity} is introduced by choosing $b_i = 0.006+0.002\cdot( (i-1) \bmod 5)$.
Since the dynamics of every unit of our network has two degrees of freedom, the system's phase space has \(D=2N\) dimensions.

The adjacency matrix~$A$ defines a chain-like topology in which units are coupled to five neighbors to their left and to five neighbors to their right if these neighbors are present, i.e.:
\begin{equation}
 A_{ij} = A_{ji} = \begin{cases} 1 & \text{for}\quad 0 < |i-j|\leq 5,\\ 0 & \text{otherwise}.
\end{cases}
\label{eq:adjmatrix}
\end{equation}
This introduces \emph{structural heterogeneities} because units at the border have a different coupling neighborhood than units in the middle of the chain-like topology. In the following, we arbitrarily chose $N=100$, but 
because of the chain-like topology defined by the adjacency matrix, we expect equivalent results for other choices of $N$. We convinced ourselves that this is indeed the case for $N=150$.

Initial conditions were chosen randomly, and we did not observe an influence of the choice of initial conditions (near the attractor) on our observations.
The equations of motion were integrated using the Runge--Kutta--Fehlberg method with a step size adapted such that the estimated relative error did not exceed $10^{-5}$ (implemented by the software package Conedy~\cite{Rothkegel2012}).
Time series were sampled with a rate of~\(1\).
To ensure that transients died out, data for the first $10^5$ time units was discarded.

\begin{figure}
\begin{center}
 \includegraphics[width=7.5cm]{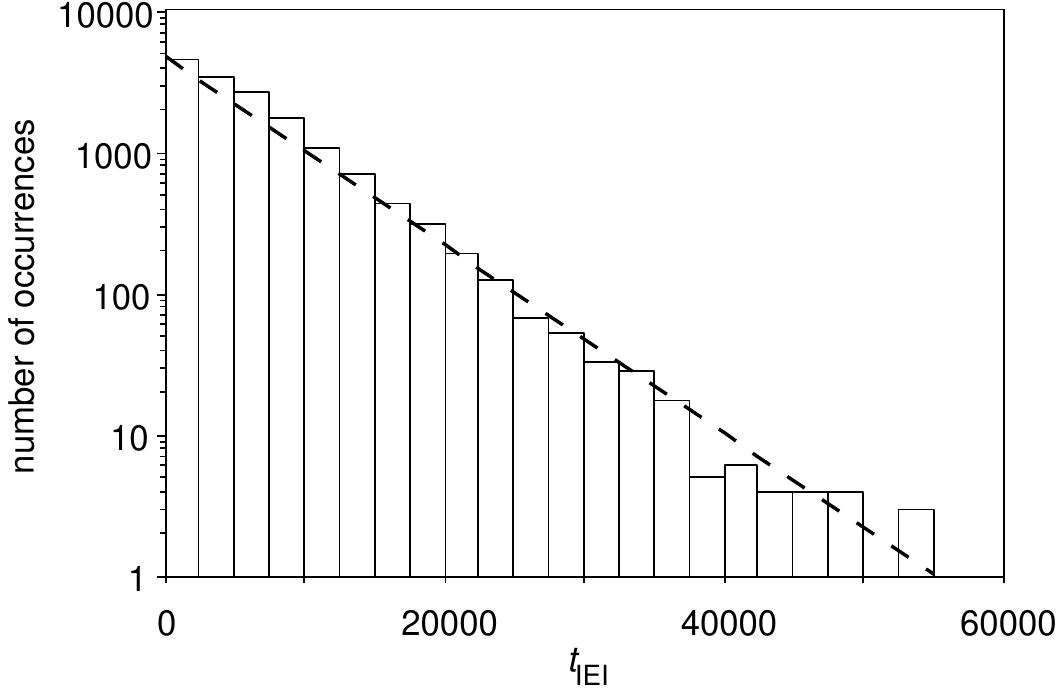}
\end{center}
\caption{
Histogram of interevent intervals $t_\text{IEI}$ obtained for a realization of the system as defined in Sec.~\ref{sec:system} (observation time of the data: $10^8$ time units). 
The rate of extreme events was estimated from the data as $r=1.54\times 10^{-4}$ (extreme events were required to be separated by at least 150 time units to qualify as separate events). 
The dashed line represents a multiple of $\exp{(-rt_\text{IEI})}$.
}
\label{fig:2}
\end{figure}

For the subsequent steps of analysis, we assume that we only observed the temporal evolution of the voltage-like variables $x_i$, $i\in\{1,\ldots,N\}$, of the system, thereby mimicking typical field studies in which not all degrees of freedom can usually be observed.
Furthermore, to ease notation, we will write~\mxc{i}{t} to refer to the value of~$x_i$ at a particular time~$t$.
In Fig.~\ref{fig:1}, we show exemplary temporal evolutions of the variables~$x_i$ (color-coded, cf.\ panel~b) for all units as well as their mean value~$\bar{x}$ (panel~a).
Most of the time, low-amplitude oscillations (panel~d) with an average period of 68~time units can be observed.
But now and then, seemingly out of a sudden, an excitation arises and travels through the network topology.
Excitations can start at different units (for example, see panels~c and~e) and thus events can take on different shapes.
The estimated distribution of inter-event intervals (cf. Fig.~\ref{fig:2}) resembles an exponential distribution (similar to the systems investigated in reference~\cite{Ansmann2013}), which would be expected for a Poissonian process, i.e.\ independently occurring events.
Since such events represent large deviations from the average dynamical behavior of the system and since they occur rarely but recurrently, we will refer to them as extreme events in the following.

\section{Construction of local embedding spaces}
\label{sec:embspaces}

The term \emph{embedding space} goes back to the seminal work of F.~Takens~\cite{Takens1981} and has become the key element of nonlinear time-series-analysis methods~\cite{Kantz2003}:
Consider a dynamical system in continuous time governed by an ordinary differential equation (ODE) with a $D$\nobreakdash-dimensional phase space and a trajectory~$\mathbf{g}(t)$ of its dynamics, emerging from an initial condition~$\mathbf{g}(0)$.
Because of the existence and uniqueness of the solution of the ODE, there exist unique mappings $\mathbf{g}(t)\mapsto \mathbf{g}(t')$ for arbitrary $t'>t$ which can be used in order to predict future values:
Knowing a state in the past at time~$t$ directly determines the future at $t' > t$.

However, in empirical studies, we are usually unable to observe the trajectory in continuous time in its full $D$-dimensional phase space.
Instead, at discrete and equidistant times $t\in\{1,\ldots,T\}$, some physical observable $u_t = h(\mathbf{g}(t))$ is recorded, where the observation function $h(\mathbf{g})$ is a scalar field on the phase space.
Given this reduced information, there is usually no unique map which maps the last observation~$u_T$ onto the future value $u_{T+1}$.

As first pointed out by Packard and coworkers~\cite{Packard1980}, the information missing in the current observation $u_T$ about the state vector $\mathbf{g}(T)$ of the dynamical system can be recovered by taking past observations into account.
Indeed, Takens' delay-embedding theorem~\cite{Takens1981,Sauer1991} states that vectors composed of successive observations, $\mathbf{u}_T=(u_T, u_{T-1},\ldots, u_{T-m+1})$ contain all information about the unknown state vector $\mathbf{g}(T)$ if the embedding dimension $m$ is larger than $2D_f$, where $D_f\le D$ is the fractal dimension of the invariant set on which the dynamics lives.
Hence, there exists a unique map $\mathbf{u}_T\mapsto u_{T+1}$ which needs to be identified from recorded data by prediction methods using sufficiently many \emph{learning pairs} $(\mathbf{u_t},u_{t+1})$.
All known methods implicitly or explicitly rely on the Lorenz method of analogues~\cite{Lorenz1969} in meteorology:
If a system's state vector comes close to a point in phase space where it had been before, then its near future will be similar to the trajectory emerging from this formerly visited point.
This is a simple consequence of the smoothness of the right-hand side of the ODE.
Thus, a simple scheme to predict $u_{T+1}$ given $\mathbf{u}_T$ is to identify the nearest neighbor $\mathbf{u}^*_t$ of $\mathbf{u}_T$ in embedding space and to set $u_{T+1} = u^*_{t+1}$.

Past attempts of such a prediction program usually aimed at predicting low-dimensional dynamics and relied on reconstructing an embedding space based on time series of one observable only.
However, as more data of spatially extended systems (e.g., gathered from distributed sensors) of large system size becomes available, the question arises how measurements of observables from different spatial locations can be exploited for prediction tasks.
For instance, when excitations propagate through a system, can measurements from neighboring locations be helpful to forecast the occurrence of an excitation at an adjacent site?

As pointed out in earlier work, components of embedding vectors can be chosen from different time series (and thus different locations of a system) in order to reconstruct an embedding space (see, e.g., Refs.~\cite{Packard1980,Kaneko1989}).
It was Parlitz and Merkwirth~\cite{Parlitz2000}, however, who assumed that local states exist in a deterministic sense which can be reconstructed in local embedding spaces:
In a spatially extended system, where the different degrees of freedom refer to different places in space, the coupling between those degrees of freedom is typically very much restricted.
As an example, consider a one-dimensional lattice and a nearest neighbor coupling, and denote by~$\mathbf{g}_i$ the degrees of freedom of a lattice site~$i$ of the phase-space vector~$\mathbf{g}$.
The near future of the site~$i$ will essentially only depend on the state of the site~$i$ itself,~$\mathbf{g}_i$, and the states of its two nearest neighbor, $\mathbf{g}_{i+1}$ and $\mathbf{g}_{i-1}$.
Hence, a local embedding aims at reconstructing the information contained in these three vectors from the array of observations, yielding local embedding spaces whose dimension can be much lower than that of the whole system's original state space and even lower than the fractal dimension as demonstrated in Ref.~\cite{Parlitz2000}.
Since for finite~$T$ these lower dimensional spaces can be populated more densely with embedding vectors, we are more likely to find good analogues that allow us to predict the next observation at site~$i$.
Doing this in parallel for every~$i$ yields a prediction of the observation array at the next time step.

\begin{figure}
\begin{center}
  \includegraphics[width=8.6cm]{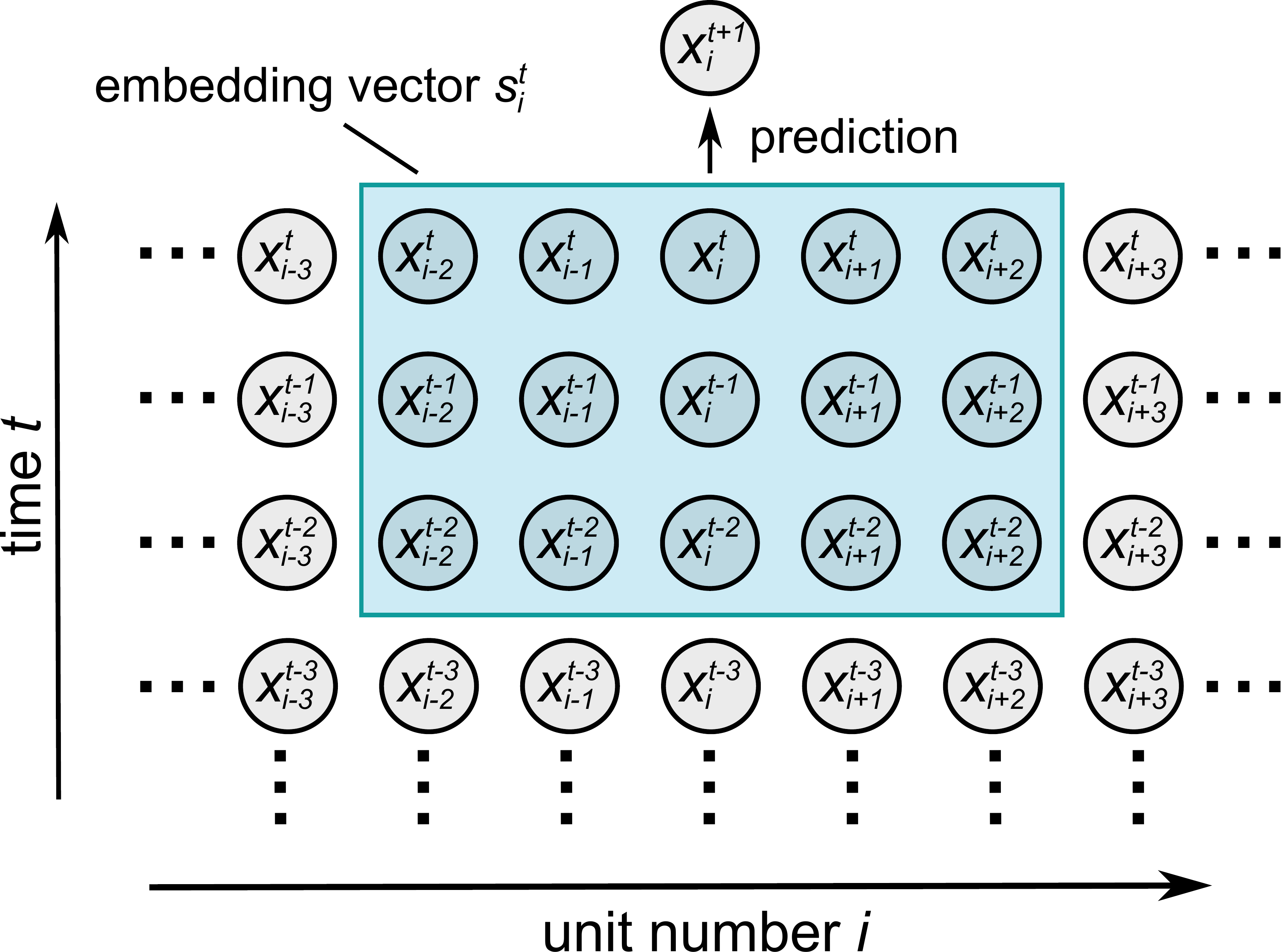}
\end{center}
\caption{(Color Online). Scheme demonstrating the construction of an exemplary embedding vector \emv{i}{t} (blue box) for unit~$i$ at time step~$t$.
The components of each embedding vector of unit $i$ are values of the multivariate time series \mx.
These values reflect the present and immediate past of unit $i$ ($t,\ldots,t-m_t$) as well as the corresponding time series values of those units $j$ which are spatially close to unit $i$ ($j\in\{i-m_s,\ldots,i-1,i+1,\ldots,i+m_s\}$ if $i$ is not close to a boundary).
The local embedding space of unit $i$ is populated with vectors $\emv{i}{T},\ldots,\emv{i}{m_t+1}$ before the future value \mxc{i}{t+1} is predicted via nearest-neighbor search.
The embedding parameters, i.e.\ the numbers of spatial neighbors $m_s$ and delayed values $m_t$, are determined by minimizing the prediction error (see text).
}
\label{fig:3}
\end{figure}

\begin{figure*}
\begin{center}
 \includegraphics{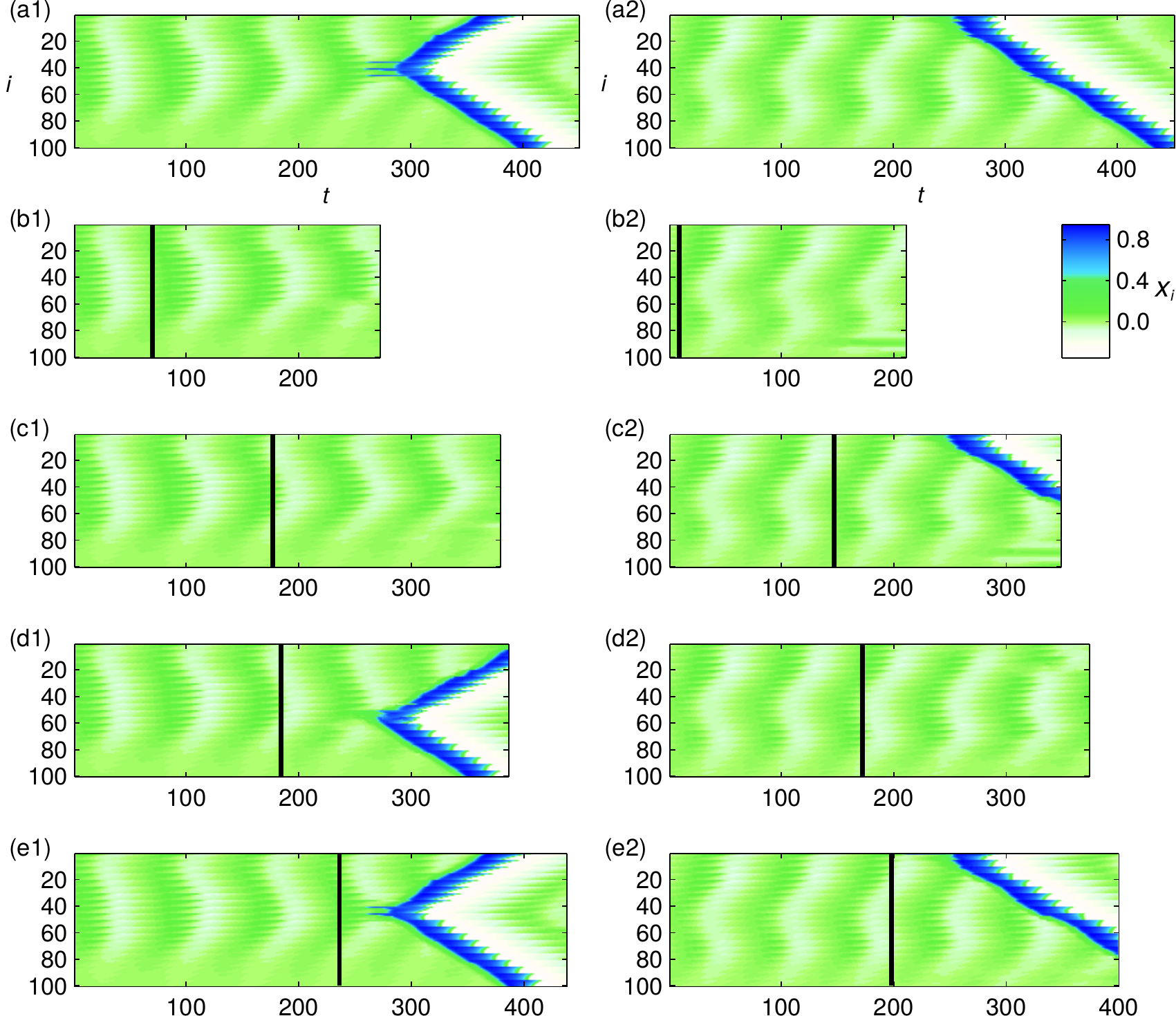}
\end{center}
\caption{(Color online) (a): Temporal evolution of the first dynamical variables $x_i$ (color coded) of all units before and during two exemplary extreme events.
(b)--(e): Predictions which consist of 200 iterative steps and start at different points in time (indicated by vertical black lines).}
\label{fig:4}
\end{figure*}

We adopted and modified this concept in order to predict extreme events in the dynamics of the system described in section~\ref{sec:system}:
In contrast to Ref.~\cite{Parlitz2000}, in which systems were homogenous and a single embedding space was used for predictions, our system is heterogeneous (parameter heterogeneity, cf.\ equation \eqref{eq:system}; structural heterogeneities, cf.\ equation \eqref{eq:adjmatrix}).
Moreover, most units are coupled to more than one nearest neighbor to their left and right.
In order to account for these heterogeneities, we construct a local embedding space $E_i$ for each unit $i$ separately.
The embedding vectors \emv{i}{t} for unit~$i$ are defined (cf.\ Fig.~\ref{fig:3}) by
\begin{equation}
\begin{aligned}
 \emv{i}{t} = \Big ( &\mxc{i-m_{si}^{*}}{t},\ldots,\mxc{i}{t},\ldots,\mxc{i+m_{si}^{\dagger}}{t},\ldots\\
 & \mxc{i-m_{si}^{*}}{t-m_t},\ldots,\mxc{i}{t-m_t},\ldots,\mxc{i+m_{si}^{\dagger}}{t-m_t} \Big ), \label{eq:embvector}
\end{aligned}
\end{equation}
with
\begin{equation}
\begin{aligned}
 m_{si}^{*}    &= m_s - \Theta(m_s - i)(m_s-i+1) \\
 m_{si}^{\dagger} &= m_s - \Theta(m_s-(N-i))(m_s-(N-i)), \label{eq:boundaries}
\end{aligned}
\end{equation}
where $\Theta$ denotes the Heaviside function ($\Theta(z)=1$ if $z\geq 0$ and $\Theta(z)=0$ everywhere else).
$m_t$~denotes the number of delayed values and $m_s$~denotes the number of spatial neighbors to the left and to the right sides of those units~$i$ that are not close to the boundaries.
Equations~\eqref{eq:boundaries} ensure that embedding vectors for units near the boundaries do only include the available spatial neighbors.
The dimensions $d_i$ of the local embedding spaces are given by
\begin{equation}
 d_i = (m_{si}^{*} + m_{si}^{\dagger} +1 )(m_t+1),
 \label{eq:dimlocem}
\end{equation}
which is reduced to $(2m_s+1)(m_t+1)$ for units not close to the boundaries (i.e.\ for $i\in\{m_s+1,\ldots,N-m_s\}$).
Given a time series of length~$T$, the embedding space~$E_i$ can be populated with $(T-m_t)$~vectors ($t\in\{T,T-1,\ldots,m_t+1\}$).

Determining the embedding parameters $m_t$ and~$m_s$ is crucial for the construction of local embedding spaces and thus for a successful prediction of the dynamics.
We constructed local embedding spaces for all combinations of values $(m_t,m_s) \in\{0,\ldots,3\} \times\{0,\ldots,7\} \backslash \{(0,0)\}$.
For each combination of values, local embedding spaces were populated with vectors derived from a training set (see section~\ref{sec:prediction}), and $N_{\text{steps}} = 40$ iterative prediction steps were undertaken to predict the beginning of a propagation of an extreme event (starting at time~$t_s$) in a test set (see section~\ref{sec:prediction}).
Let \mxc{i}{t} denote the value of the excitatory variable of unit~$i$ at time~$t$, and let \mxct{i}{t} denote the corresponding value that was predicted by iterative predictions based on local embedding spaces with parameters $m_t$ and~$m_s$.
We used the root of the mean squared error $\psi$, defined as
\begin{equation}
 \psi_{m_t,m_s} = \sqrt{ \frac{\sum_{i=1}^{N}\sum_{t = t_s}^{t_s+N_{\text{steps}}} \left (\mxct{i}{t} - \mxc{i}{t}\right )^2}{NN_{\text{steps}}}},\label{eq:rmse}
\end{equation}
to quantify the prediction error \cite{Farmer1987,Casdagli1992}.
We observed $(m_t,m_s)=(2,2)$ to yield the lowest $\psi$ and used these embedding parameters in all subsequent prediction tasks.
This choice led to local embedding spaces with a maximum dimension of~15 (cf.\ equation~\eqref{eq:dimlocem}).

\section{Data-based prediction of extreme events}
\label{sec:prediction}

\begin{figure*}
\begin{center}
 \includegraphics[width=\textwidth]{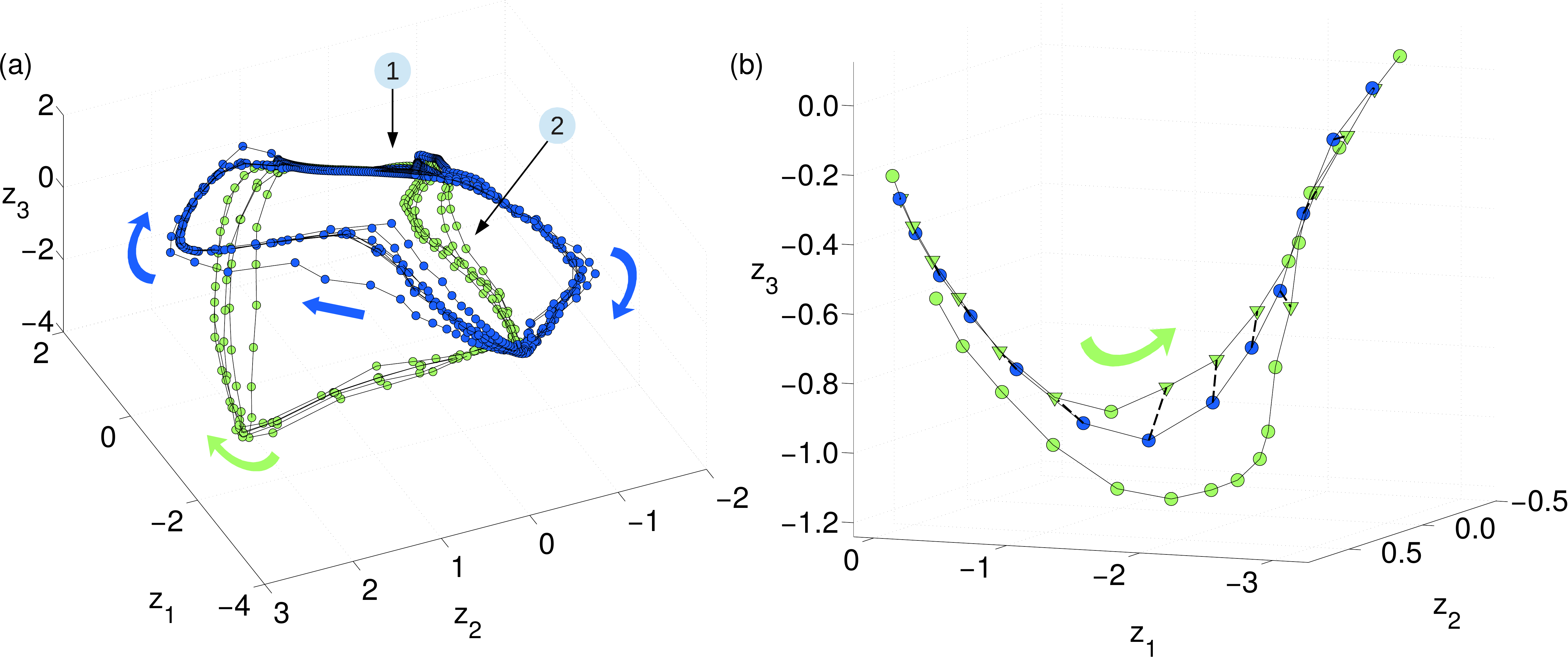}
\end{center}
\caption{(Color online) (a)~Three-dimensional projection of a 15-dimensional local embedding space of unit~40 constructed from time series of the training set (which included 13 extreme events).
(1)~marks a small subspace in which all vectors representing low-amplitude oscillations reside.
Vectors corresponding to excitations (i.e.\ extreme events) propagating over unit~40 from lower-numbered units are colored blue, those from higher-numbered units are colored green.
(b) Enlargement of region~(2) of panel~(a) showing vectors of the training set (green) as well as an iterative prediction (blue) representing an excitation of unit~40.
Each vector from the last prediction step is connected to its nearest neighbor (green triangle) with a dashed line.
The image of the nearest neighbor as well as the vectors of the neighboring units are used to construct the next predicted vector.
Colored arrows in both panels indicate the direction of temporal evolution.
The axes $z_1$, $z_2$, $z_3$ span the space of the three-dimensional projection as determined by isomap \cite{Tenenbaum2000, Lee2007}, using $\epsilon_\text{NG}=0.5$ for the construction of the neighborhood graph.
}
\label{fig:5}
\end{figure*}

Time series of the voltage-like variables \mxc{i}{t}, $i\in\{1,\ldots,N\}$, of the system dynamics were obtained by integrating the equation of motions as detailed in section~\ref{sec:system}.
For each unit~$i$, a local embedding space was populated with vectors constructed from the multivariate time series (training set) of $T=10^5$ data points, containing 13~extreme events of different shape.
The system was further integrated with the same parameters to obtain a test set containing 66~extreme events.
Embedding parameters of the local embedding spaces were chosen according to the previous section.

Knowing the last observations~$\mathbf{x}^{t}$ of the system, we predict the future values $\mathbf{\tilde{x}}^{t+1}$ by making use of Lorenz method of analogues~\cite{Lorenz1969}.
For each unit $i$, we construct the vector~\sit{i}{t} that corresponds to~\mxc{i}{t} in the embedding space~$E_i$.
The vector~\sit{i}{t'} of the training set with the smallest Euclidean distance to~\sit{i}{t} (i.e.\ the nearest neighbor of~\sit{i}{t}) is determined.
The vector \sit{i}{t'+1} is obtained and its associated value \mxc{i}{t'+1} is taken as the predicted value, i.e.\ $\tilde{x}_i^{t+1} = \mxc{i}{t'+1}$.
After predictions have been obtained for all $i$, $\mathbf{\tilde{x}}^{t+1}$ are used to construct a new embedding vector in each of the $N$ embedding spaces.
The future values $\mathbf{\tilde{x}}^{t+2}$ can then be predicted by applying the same prediction scheme but now starting with $\mathbf{\tilde{x}}^{t+1}$.
This way, iterative predictions can be made arbitrarily far into the future.

In Fig.~\ref{fig:4} we show the temporal evolution of two exemplary extreme events (panels (a1) and~(a2)) from the test set as well as attempts to predict these events (panels (b1)--(e1) and (b2)--(e2)).
The period of time between the start of the iterative predictions (indicated by black vertical lines) and the onset of the actual extreme event (here defined as the first point~$t^*$ in time for which $\mxc{i}{t^*}>0.22$ for some~$i$) is called the lead time.
Each prediction attempt consisted of 200 iterative prediction steps.
For lead times larger than 200, we observe the method to correctly predict low-amplitude oscillations of all units (panels (b1) and~(b2)).
However, as the iterative predictions proceed, prediction errors accumulate which is reflected in small artifacts (cf.\ panel~(b2) for $t\in [150,210]$ and $i\in[85,95]$).
For the first event, when the lead time is decreased, low-amplitude oscillations are predicted and the extreme event is missed (panel~(c1)).
For a smaller lead time (panel~(d1)), the method predicts low-amplitude oscillations, followed by the onset as well as the propagation of an extreme event, which does not precisely coincide with the spatiotemporal onset of the event in the original dynamics, though.
Finally, when decreasing the lead time even further (panel~(e1)), we observe the predicted onset as well as the propagation of the extreme event to closely reflect the event in the original dynamics.

We observed this scheme -- better predictability for smaller lead times -- to hold for many prediction attempts and we will quantify this observation in the next section.
However, we also observed a sensitive dependence on the lead time:
For instance, for the second event depicted in Fig.~\ref{fig:4}, decreasing the lead time first led to a successful prediction of an extreme event (panel (c2)), but for a further decrease, our method missed to predict the event (panel~(d2)).
Only after decreasing the lead time even further, the extreme event was again successfully predicted (panel~(e2)).
We hypothesize that this observation is related to different nearest neighbors found in embedding spaces depending on the lead time.

In order to get more insight into the structures present in the local embedding spaces, we show in Fig.~\ref{fig:5}~(a) a three-dimensional projection of the 15-dimensional embedding space of unit~40 with vectors from the training set.
In region~(1) of panel~(a), we observe a limit-cycle-like structure (depicted from the side), which reflects the dynamics associated with low-amplitude oscillations.
From this structure, two bunches of trajectories emerge, which reflect the excitations of unit~40 during extreme events and can be distinguished with respect to the direction in which excitations propagate (blue vs. green).
Thus, local features of the dynamics of this unit being embedded in a larger network are reflected by embedding spaces.

In Fig.~\ref{fig:5}~(b), we show a zoom-in of region~(2) of panel~(a) with additional vectors (blue dots) that were obtained by a prediction of the dynamics of unit~40 for an extreme event from the test set.
The predicted trajectory evolves along existing trajectories in embedding space but differs from them.
This is due to the construction of embedding vectors:
Each (predicted) embedding vector incorporates not only information about the temporal past of the unit but also information from topological neighbors.

\section{Statistical results on forecast performance}
\label{sec:statistics}
The extreme events in our system start with an excitation of one or a few units, after which the excitation propagates through the network topology.
To evaluate the forecast performance of our method, we focus on the prediction of the onset of extreme events.
The spatiotemporal onset is defined as in the previous section, i.e.\ it is the point in time~$t^*$ and the set of units~$i$ whose excitatory variables~$x_i$ first cross a predefined threshold ($0.22$).
Since Fig.~\ref{fig:4} suggests that forecast performance may depend on the lead time (i.e.\ the time between the start of our prediction attempt and the actual onset of the extreme event), we quantify prediction performance depending on the lead time or, equivalently, on the number of iterative prediction steps.

For possible applications, two quantities are of particular interest: the probability to correctly predict the onset of an extreme event and the probability to (wrongly) predict an onset of an event that is not present in the original dynamics.
For practical uses, it may even be sufficient if the predicted onset (at time $t_\text{pred}$ and unit $i_\text{pred}$) is spatiotemporally close to the actual onset (at time $t_\text{true}$ and unit $i_\text{true}$).
We thus require that a correct prediction of an onset needs to satisfy $|t_\text{pred}-t_\text{true}| < \epsilon_t$ and $|i_\text{pred}-i_\text{true}| < \epsilon_s$.
Increasing $\epsilon_s$ and $\epsilon_t$ will make correct predictions of onsets easier.
In the rare cases in which predicted onsets or actual onsets consist of more than one unit, we require that our method correctly predicts at least one unit involved in the onset.

To quantify the probability to correctly predict onsets of extreme events (i.e.\ the true-positive rate, TPR), an ensemble of $P_e = 66$ extreme events was selected from the test set.
We chose a lead time $\tlead$ and, for each event in the test set, we started a prediction attempt (consisting of 200 iterative prediction steps) at that lead time.
The true-positive rate is then defined as $\text{TPR}(\tlead)= \text{TP}(\tlead) / P_e$, where $\text{TP}(\tlead)$ is the number of correct predictions (according to the previous paragraph) for lead time \tlead in test set.

Prediction errors likely accumulate with increasing number of iterative prediction steps, which may lead to the prediction of extreme events which are not present in the actual dynamics (false positives).
Several approaches can be conceived to quantify the probability of obtaining false positives.
One approach would be to estimate the probability of obtaining a false positive for a particular step \tstep within the iterative prediction steps.
However, in practical applications it is usually more informative to learn about the probability to erroneously predict an extreme event within the first \tstep iterative prediction steps.
To quantify the latter, we regarded an ensemble of $N_e=1000$ randomly chosen, non-overlapping and event-free intervals (each of length 200 time units) from the test set.
We started a prediction attempt (consisting of 200 iterative steps) at the beginning of each interval.
For each attempt $e\in\{1,\ldots,N_e\}$, we determined the iterative step~$t_e^*$ where an onset of an extreme event was (erroneously) predicted.
If no onset was predicted, we set $t_e^*=\infty$.
The probability of predicting an extreme event not existing in the original dynamics within the first \tstep iterative prediction steps (i.e.\ the false-positive rate, FPR) is then defined as $\text{FPR}(\tstep) = N_e^{-1} \sum_{e=1}^{N_e} \Theta(\tstep - t_e^*)$.

\begin{figure}
\begin{center}
 \includegraphics[width=7.5cm]{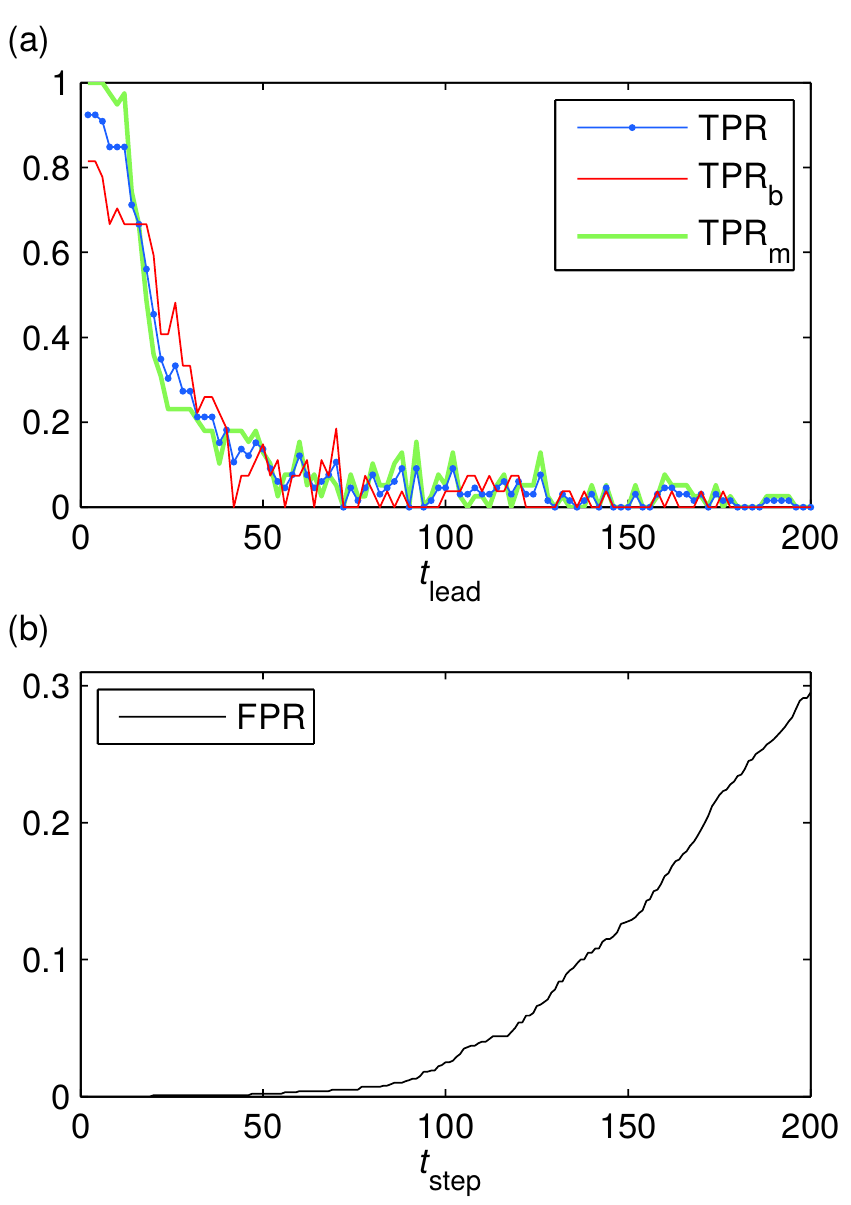}
\end{center}
\caption{(Color online) (a) Dependence of the true-positive rate on the lead time \tlead determined from all extreme events in the test set (TPR, dotted blue line, 66~events), determined only from the extreme events with onsets at the border of the chain-like topology ($\text{TPR}_\text{b}$, red, 27~events), and from those with onsets everywhere else ($\text{TPR}_\text{m}$, green, 39~events).
TPR values were obtained using $\epsilon_t=3$ and $\epsilon_s=5$.
(b) Dependence of the false-positive rate (FPR) on the number of iterative prediction steps \tstep.}
\label{fig:6}
\end{figure}

Figure~\ref{fig:6}~(a) shows the dependence of the true-positive rate (determined for all $P_e$~events in the test set) on the lead time.
We observe the TPR to (roughly) increase for decreasing lead time which is in accordance with our observations of the previous section that events become better predictable in space and in time the closer they are in time.
The TPR decreases fast with increasing lead time, crossing the $0.5$ value at $\tlead\approx 19$.
However, we still found TPR values above $0.1$ for larger lead times (e.g.\ $\text{TPR}(70)=0.11$).
For small lead times ($\tlead < 14$), we obtained TPR values larger than $0.8$ but did not observe values of~1, i.e.\ a true-positive rate of 100\,\%. We determined the true-positive rate for various values of $\epsilon_s$ and $\epsilon_t$ (data not shown) and observed the TPR only to slightly increase with $\epsilon_s$ and $\epsilon_t$ (depending on the lead time), indicating that if our method predicted an onset, it was in the spatiotemporal vicinity of the true onset.

To investigate whether the predictability of extreme events differed according to onset locations, we divided the ensemble of extreme events into two ensembles, one containing events with onsets at the border of the chain-like topology ($\text{TPR}_\text{b}$ in Fig.~\ref{fig:6}~(a), red line) and one containing all other ones ($\text{TPR}_\text{m}$, green line).
We observed the TPR for extreme events of the latter group to obtain larger values for small lead times ($\text{TPR}_\text{m} = 1$ for $\tlead<8$) than for those in the former group (maximum value: $\text{TPR}_\text{b} = 0.81$).
We investigated whether this difference in predictability may be attributed to properties of the training set. 
As the quality of the prediction of the dynamics depends on the ability to find good analogues in local embedding spaces (i.e.\ nearest neighbors which are very close to the last known state), we investigated at which units we could observe single excitations to start and later to spread to neighboring unit. 
We found that the training set contained such a pattern for every unit that was involved in an onset in the test set, including the one at the border (4 out of the 13 extreme events in the training set started at the border).
Thus the difference in predictability of extreme events starting at the border cannot be explained by a lack of corresponding analogues in the training set.

The false-positive rate (FPR) is below 1\,\% for $\tstep<86$ (cf.\ Fig.~\ref{fig:6}~(b)).
As the number of iterative steps increases, prediction errors accumulate leading to an increase of the FPR.
For the maximum number of iterative prediction steps investigated here (200 steps), in 30\,\% of prediction attempts an extreme event is predicted to occur which does not occur in the actual dynamics ($\text{FPR}(200)=0.3$).

Prediction attempts aiming at predicting extreme events long in advance (which implies large lead times) will produce many false positive findings (large FPR for a large number of iterative steps) whereas the probability of a correct prediction is low (low TPR).
On the other hand, prediction attempts with small lead times come along with high probabilities to correctly predict the occurrence of an extreme event (large TPR for small lead times) and with low probabilities of false alarms (low FPR), but provide only short time windows for possible interventions or warnings. 
Thus, we face a tradeoff between a desired large lead time on the one side and desired optimal values for TPR and FPR on the other. 
We note that we do not face a tradeoff between TPR and FPR given the lead time as a free parameter: 
Decreasing the lead time optimizes both (i.e.\ TPR increases while FPR decreases) which is the reason why we refrain from conducting a receiver operating characteristic (ROC) analysis~\cite{Fawcett2006} (which is suited for situations in which increasing TPR values come along with increasing FPR values).

\section{Preventing extreme events with small interventions}
\label{sec:prevention}

The ability to forecast extreme events in our system can provide us with a period of time during which an event is predicted to occur but has not yet started.
During such time spans, interventions (i.e.\ perturbations of the system dynamics) could be applied to prevent extreme events.
We investigated whether small perturbations of single units at single points in time can prevent extreme events from occurring in our system and when and where such perturbations are most effective.
To this end, we selected an extreme event from the test set.
For each series of perturbation experiments, we chose a time point~$t_\text{p}$ and a unit~$i$ and added a small positive value (i.e.\ the perturbation) to the variable $y_i(t_\text{p})$.
We integrated the equations of motions and evaluated whether an extreme event occurred within a period of time after the perturbation.
During the evaluation period, which started at $t_\text{p}$ and ended $98$~time units after the onset in the unperturbed dynamics, we checked whether $\bar{x}$ exceeded~$0.1$ (which indicated the occurrences of an extreme event; cf.\ Fig.~\ref{fig:1}~(a)).
Repeating this experiment with varying amplitude, we identified for each~$i$ and~$t_\text{p}$ the minimum perturbation amplitude that was sufficient to prevent the extreme event from occurring.

\begin{figure}
\begin{center}
 \includegraphics[width=8.6cm]{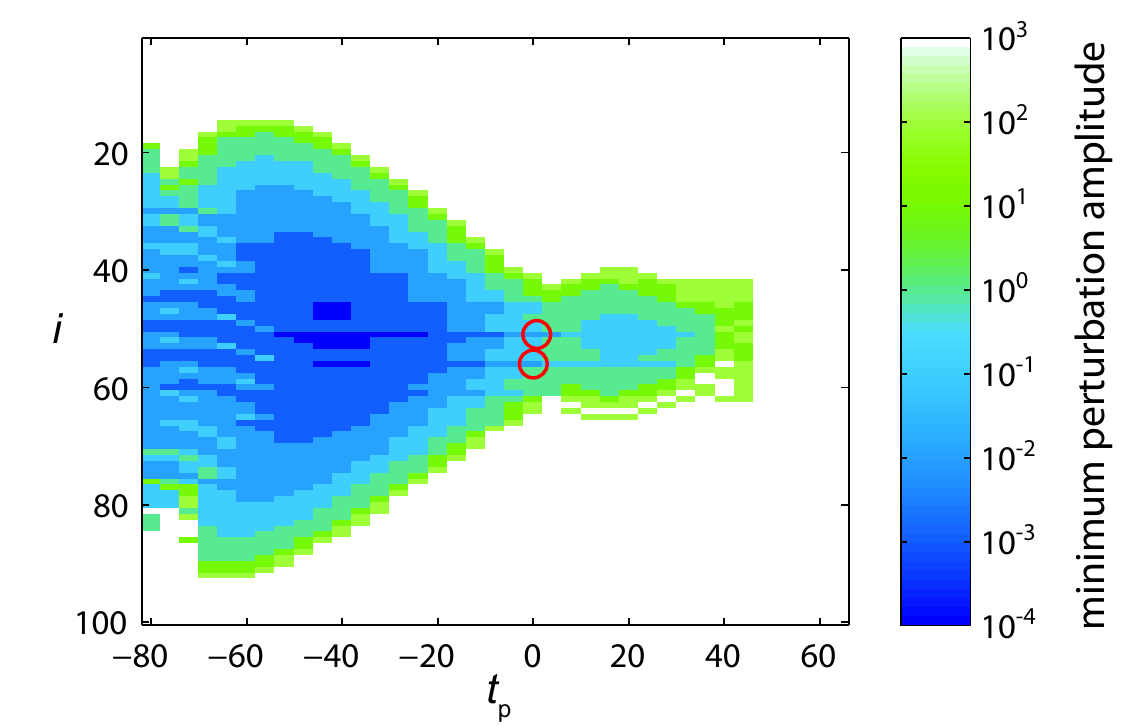}
\end{center}
\caption{(Color online) Minimum amplitude (color coded, capped at $10^3$) of a perturbation at time~$t=t_\text{p}$ of the inhibitory variable of unit~$i$ to prevent an exemplary extreme event from occurring before $t=98$.
The onset of the extreme event in the case of unperturbed dynamics was located at $t=0$ and $i=56$ (indicated by a red circle).
At $t=1$, a second unit, $i=51$ (red circle), crossed the threshold.}
\label{fig:7}
\end{figure}

In Fig.~\ref{fig:7}, this amplitude is shown for different units~$i$ and times~$t_\text{p}$ and for an exemplary event.
We observe the minimum perturbation amplitudes to be orders of magnitudes smaller for units involved in or topologically close to the onset (e.g.\ $10^{-4}$ for units $51$ and $56$ at $t_\text{p}=-40$) than for those more distant to the onset (e.g.\ $10^{2}$ for unit $i=19$ at $t_\text{p}=-40$).
For units that are topologically far away from the onset, the maximum perturbation amplitude ($10^3$) investigated in our experiments was not sufficient to prevent extreme event from occurring (e.g.\ units $i<15$ and $i>92$).
Moreover, we observed the region of units with small perturbation amplitudes ($<10^{-1}$) to decrease when approaching the onset of the extreme event.
Even in the time span after the onset but before the spreading of the excitation to neighboring units ($0\le t_\text{p}\le 45$), the extreme event could still be prevented by perturbing single units.
However, in this case, the necessary perturbation amplitudes are orders of magnitudes larger than those for the same units before the onset.
Thus, predicting the onset of an extreme event enables us to prevent it via a minimum intervention, i.e.\ by applying a perturbation of minimum amplitude.

Combining the insights gathered in the previous experiments, we investigate whether repeated predictions of the dynamics together with small perturbations can control the dynamics such that no extreme events occur any more.
Initial conditions were randomly chosen, and the dynamics was integrated for 10 time units, after which the future temporal evolution was iteratively predicted (using the same embedding spaces as in section~\ref{sec:prediction}).
If an onset was predicted, a perturbation with amplitude $10^{-1}$ was applied to the unit of the event's onset.
If the forecast predicted the voltage-like variable of other units to also cross the threshold within five time units after the onset, a perturbation was also applied to up to 2 of these additional units.
Whenever a perturbation was applied, we also integrated the unperturbed dynamics of the system in order to assess whether the prediction of an onset of an extreme event was correct (true positive) or not (false positive).
If no onset was predicted, no perturbation was applied.
Afterwards and in both cases, the dynamics was integrated for another 10~time units.
The steps of prediction, possible intervention, and integration were repeated until a total time span of 204,000 time units was obtained.

\begin{figure*}
\begin{center}
 \includegraphics[width=17.8cm]{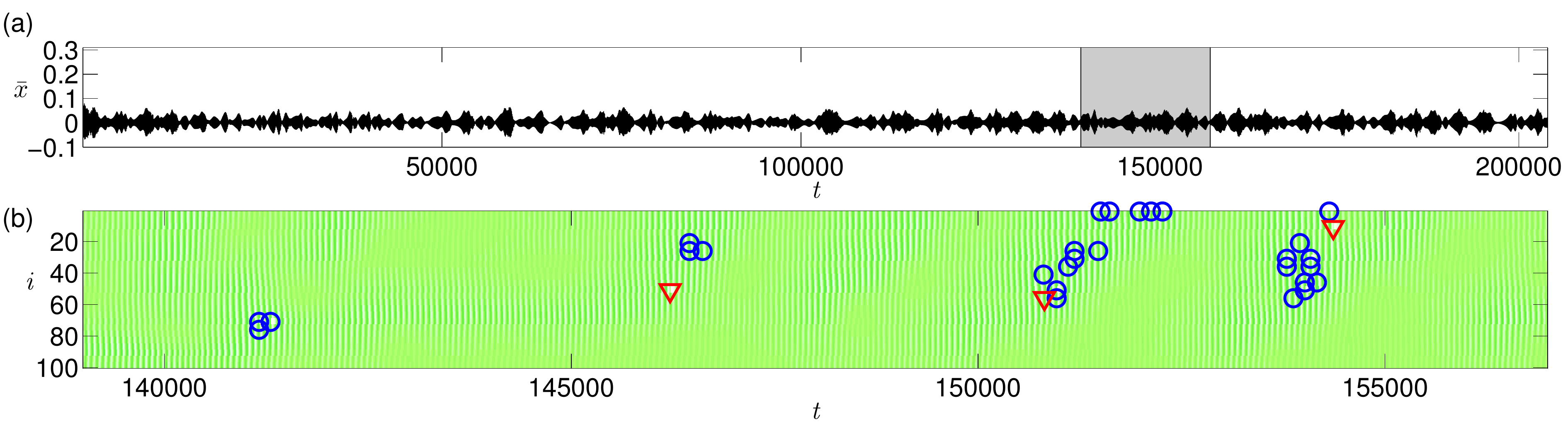}
\end{center}
\caption{(Color online) (a)~Temporal evolution of the mean value $\bar{x}$ of the voltage-like variables $x_i$, $i\in\{1,\ldots,N\}$ for a control experiment spanning 204,000 time units.
The gray background marks the region for which the temporal evolution of all $x_i$ is shown in panel (b).
(b)~Temporal evolution of $x_i$ (color coded as in Fig.~\ref{fig:1}b) for all units $i$ and for an exemplary excerpt of panel~(a) (marked gray).
Blue circles mark points in time and in space for which prediction attempts correctly predicted the occurrence of an extreme event.
Red triangles mark points in time and in space for which prediction attempts erroneously predicted the occurrence of an extreme event.}
\label{fig:8}
\end{figure*}

For our iterative predictions, we chose a lead time of 30 time units because for this lead time the probability of false positives is low ($\text{FPR}(30)=10^{-3}$, cf.\ Fig.~\ref{fig:6}~(b)).
At this lead time, the probability of correctly predicting the onset of an extreme event is non-zero but still small ($\text{TPR}(30)=0.26$).
However, if the prediction scheme misses to predict an extreme event for large lead times ($\tlead > 20$), then the extreme event can still be located after the next integration step with a larger probability.
Due to this strategy, there are up to three opportunities (coming with different lead times) for our prediction scheme to forecast a given event.

In panel (a) of Fig.~\ref{fig:8}, we illustrate the effect of this prevention scheme.
The mean value $\bar{x}$ never exceeds $0.1$, which shows that no extreme event occurred during the simulation (cf.\ section~\ref{sec:system} and Fig.~\ref{fig:1}~(a)).
We observed that onsets of extreme events were predicted (cf.\ panel~(b)), and interventions were carried out.
Among the total number of 262 predicted onsets in the simulation, 188 (72\,\%) were correct (true positives; blue circles) and 74 (28\,\%) were erroneous (false positives; red triangles).
Together with 20401 total prediction attempts during the simulation, this yields a small false-positive rate of $3.6\times 10^{-3}$.
For the unperturbed dynamics of the simulation period, the expected number of extreme events is 31 (given an extreme events rate of $1.54\times 10^{-4}$; cf. Fig.~\ref{fig:2}), which is smaller than the number of true positives.
This can be explained by the observation that small perturbations sometimes prevent events by ``delaying'' them, yielding predicted onsets which are clustered in time (cf. Fig.~\ref{fig:8} (b)).
To briefly summarize, we were able to successfully prevent extreme events via small interventions which were targeted to perturb just very few units at particular points in time.

\section{Discussion}
\label{sec:discussion}
The rareness of extreme events, which is a consequence of their impact and implied by their definition (societies and environments usually adapt to frequently occurring, and thus ordinary, events)~\cite{Kantz2006}, poses particular challenges for their prediction.
First, regular prediction attempts to forecast rare events require small false-positive rates, in particular if erroneous warnings or preventions are expensive, consume limited resources, or come along with unwanted side effects.
Our results in section~\ref{sec:prevention} reflect this requirement: despite a low false-positive rate ($\text{FPR}<0.01$ for $\tstep < 86$), 28\,\% of all positive findings were false positives (FP).
The large number of prediction attempts (due to the rareness of extreme events) translates into a large number of attempts which should raise no warning (i.e.\ a large number of negatives, $N_e$).
Since $\text{FP}=\text{FPR}\times N_e$ increases with the number of negatives (which, in turn, becomes larger the rarer the events), this calls for prediction methods with exceptionally small FPR values.
Such values come along with small lead times (cf.\ Fig.~\ref{fig:6}), limiting the warning time before an extreme event.

Second, in field studies, the rareness of extreme events implies that measurements need to be performed over a long period of time in order to observe a reasonable number of extreme events for the training set upon which the prediction method is based.
The rare occurrence of extreme events (``rare trajectories'') leads to local embedding spaces in which most vectors reflect the regular dynamics and only a few reflect the dynamics before or during an extreme event (cf.\ Fig.~\ref{fig:5}(a)).
This can pose challenges with respect to efficient nearest-neighbor searches, storage requirements and, more importantly, the ability of the method to find good analogues before an impending extreme event.
Limited storage capabilities could be more efficiently used by choosing vectors in a way that allows for a selective and denser coverage of regions of interest in embedding spaces such as those reflecting the dynamics before and during extreme events.
Such a sampling strategy, which would also relax computational requirements for efficient nearest-neighbor searches, has already been proposed in the context of data streams~\cite{Kwasniok2004}.
If different initial conditions can be explored, importance sampling Monte Carlo methods may also provide an efficient means towards a denser coverage of regions of interest in embedding spaces~\cite{*[][{ and references therein}] Leitao2014}.

Prediction performance may be improved in different ways.
We observed our method to yield different true-positive rates depending on where the onsets of extreme events were located (cf.\ Fig.~\ref{fig:6}(a)).
This finding may suggest that some extreme events observed in our dynamics might be intrinsically better predictable than others.
However, the differing true-positive rates may also be related to the way how embedding parameters were determined.
For all units, the same embedding parameters $(m_t,m_s)$ were determined by minimizing the root-mean-square error globally (cf.\ equation~\eqref{eq:rmse}), leading to local embedding spaces of different dimensionality (e.g.\ $d_1=d_{100}=9$ at the border and $d_{50} = 15$ in the center of the chain-like topology; cf.\ equation~\eqref{eq:dimlocem}).
Thus, the true-positive rate may be improved by determining embedding parameters for each unit separately, e.g.\ by minimizing a local root-mean-square prediction error.

Further improvements in prediction performance may be achieved by devising methods for the construction of embedding vectors which systematically account for physical scales present in the dynamics, for instance, by introducing varying and adjusted lags between consecutive vector components~\cite{Pecora2007} or by choosing a different spatiotemporal shape for the local embedding vectors: the rectangular region (cf.\ Fig.~\ref{fig:3}) could be replaced by a triangular one \cite{Mandelj2000,Mandelj2001}, a ``pyramid''~\cite{Olbrich2000}, or a ``light cone''~\cite{Parlitz2000} to account for the fact that physical information propagates at some maximum velocity.
Employing more sophisticated schemes to obtain predictions in local embedding spaces may yield further improvements in prediction performance.
For instance, the average of images of nearest neighbors may provide a more robust prediction than the image of one nearest neighbor alone; or local linear models or a global nonlinear model could be fitted to the structures in local embedding spaces to assist predictions \cite{Kantz2003}.
Besides, predictions will improve with an increasing number of vectors contained in embedding spaces as better analogues become available.
For finite multivariate time series, more densely populated embedding spaces may be obtained by exploiting symmetries of the system if they exist and are known:
If some subsets of units of the dynamics are similar with respect to their local dynamics and local coupling neighborhood, similarity classes could be defined.
The embedding space associated with a similarity class would incorporate all embedding vectors created from the data of the units of this class.
Such a strategy can be successful for homogenous systems (as demonstrated in~\cite{Parlitz2000}) and may possibly be even successful for weakly heterogeneous systems (as suggested by data-assimilation experiments \cite{Sauer2009}).

In order to define local embedding spaces, we relied on the knowledge of the coupling topology (i.e.\ the adjacency matrix) of the system.
Together with the embedding parameters, the adjacency matrix determined the neighbors to consider when constructing vectors of a local embedding space.
While in modeling studies the adjacency matrix of the system under consideration is known, in field studies this is usually not the case.
In spatially extended systems, observables which are spatially close may be considered as natural coupling candidates.
However, spatial closeness might not always be a good criterion for dynamical systems in which observables can be related to each other at large distances (e.g., teleconnections in the climate system \cite{Bridgman2006}, or long-distance pathways in the human brain \cite{Sporns2011a}).
The inference of the adjacency matrix from empirical data may be thus recognized as a first step towards a successful prediction of the dynamics for such systems.
Recent years have seen progress to this respect~\cite{*[][{ and references therein}] Timme2014} yielding methods differing according to the influence which can be exerted on the system:
If the system can be driven, driving-response based methods (see, e.g., Ref.~\cite{Yu2010}) may be appropriate, whereas correlation-based methods, which are likely affected by the spatial~\cite{Bialonski2010} and temporal~\cite{Bialonski2011b} sampling of the dynamics, may be considered for systems whose dynamics can be only observed but not manipulated in a controlled way (see, e.g., Ref.~\cite{Kramer2009}).

We prevented extreme events to occur by perturbations which were localized in space and in time according to the predicted onset of extreme events.
Alternative prevention schemes may yield even lower perturbation strengths, for instance, by expanding the period of time during which a unit is perturbed or by perturbing multiple units in the spatiotemporal vicinity of the predicted onset.
Targeted perturbations were also reported to be successful in preventing extreme events in the complex Ginzburg-Landau equation \cite{Nagy2007,Du2008,Du2011}.
In these studies, extreme events were prevented by applying a local perturbation \cite{Nagy2007,Du2008} or by perturbing a whole region \cite{Du2011} whenever an observable crossed a specified threshold (this prevention scheme was also used for low-dimensional dynamics \cite{Cavalcante2013,ZamoraMunt2014}).
We note that such a strategy could also be adopted for our system: The crossing of a threshold would indicate the beginning of an extreme event and would coincide with our definition of the spatiotemporal onset (cf.\ Fig.~\ref{fig:7}, $t_\text{p}=0$).
Figure~\ref{fig:7} illustrates that extreme events could still be prevented (for $0< t_\text{p} \leq 45$), however, using perturbation strengths which are orders of magnitudes larger than those required when applied before the threshold crossing.

We close this discussion by noting that the construction of local embedding spaces may be considered as a strategy which avoids high-dimensional spaces.
The latter often come along with the ``curse of dimensionality'', i.e.\ a finite amount of data rapidly becomes sparse as the dimension of the space (and thus its volume) increases~\cite{Lee2007}.
Moreover, with increasing dimension, Euclidean distances between any two vectors tend to become more and more similar and thus less discriminative (also known as \emph{concentration of norms),} which can cause problems for nearest-neighbor searches \cite{Beyer1998,Hinneburg2000} and time-series prediction~\cite{Verleysen2005a}.
To deal with the challenges associated with high-dimensional systems, alternative strategies may comprise dimensionality-reduction techniques applied as a pre-processing step~\cite{Verleysen2005a}, the use of alternative distance metrics~\cite{Aggarwal2001}, or of reduced models which mimic the extreme-value characteristics of the full models and thus may allow for a successful statistical prediction of extreme events~\cite{Franzke2012}.

\section{Conclusions}
We predicted the beginning, propagation, and termination of extreme events in a spatially extended, heterogeneous, excitable system, in which events occur in a temporally and spatially irregular way.
Iterative predictions of the beginning of extreme events enabled us to suppress events completely by small, rare, and spatiotemporally localized perturbations.
The predictions were based on time-series data, replicating the situation of many field studies, in which the governing equations of motion of the studied systems are typically unknown and only a subset of all degrees of freedom of the system can be observed for a finite period of time.
Our approach is based on the concept of local embedding spaces~\cite{Parlitz2000}, in which vectors incorporate information of observables of coupling neighbors, thereby featuring local characteristics of the dynamics.
Local predictions of the beginning of extreme events yielded remarkably low false-positive rates and large true-positive rates for sufficiently small lead times.

Further work will be needed before our prediction approach can be successfully applied to natural systems.
Besides theoretical work, which could shed light on the concept of local states, we consider research into the robustness of our method promising.
How will forecast performance be affected by noise contributions (introduced, e.g., by measurement uncertainties or by stochastic components in the system dynamics) or by uncertainties when constructing local embedding spaces (possibly induced when inferring the coupling topology from empirical data)?
Furthermore, the development of ideas and methods which are specifically designed to deal with ``rare trajectories'' may significantly advance our ability to forecast extreme events.

\begin{acknowledgments}
The authors thank Rajat Karnatak for helpful discussions.
SB and GA acknowledge support by the Volkswagen Foundation (Grant No.~85390 and~85392, respectively).
\end{acknowledgments}

\end{document}